\documentclass[sigplan,10pt]{acmart}
\renewcommand\footnotetextcopyrightpermission[1]{}
\AtBeginDocument{%
  }

\usepackage{multirow}
\usepackage{xparse}
\usepackage{xcolor}
\usepackage{enumitem}
\usepackage[ruled,linesnumbered]{algorithm2e}
\usepackage{mathrsfs}
\usepackage{makecell}
\usepackage{subcaption}
\usepackage{tablefootnote}
\usepackage{booktabs,tabularx}
\usepackage{amsmath}
\usepackage{graphicx}
\usepackage{adjustbox}
\usepackage{makecell}
\usepackage{caption}
\usepackage{titlesec}
\makeatletter

\titlespacing*{\section}{0pt}{6pt plus 1pt minus 1pt}{2pt}
\titlespacing*{\subsection}{0pt}{4
pt plus 1pt minus 1pt}{2pt}

\setlist[itemize]{itemsep=2pt,topsep=2pt}

\newcommand{\PP}{\textsc{RAGDoll}}

\newcounter{siqiang}
\numberwithin{siqiang}{section}
    
\NewDocumentCommand{\leo}{mg}{\IfNoValueTF{#2}
{\textcolor{teal}{#1}}
{\textcolor{teal}{$\blacktriangleright$}#1 \textcolor{teal}{$\triangleright$ #2$\blacktriangleleft$}}}

\setcopyright{acmlicensed}
\copyrightyear{2018}
\acmYear{2018}
\acmDOI{XXXXXXX.XXXXXXX}
\acmConference[Conference acronym 'XX]{Make sure to enter the correct
  conference title from your rights confirmation email}{June 03--05,
  2018}{Woodstock, NY}
\acmISBN{978-1-4503-XXXX-X/2018/06}

\begin{document}

\title{RAGDoll: Efficient Offloading-based Online RAG System on a Single GPU}

\author{Weiping Yu}
\authornote{Both authors contributed equally to this research.}
\email{weiping001@e.ntu.edu.sg}
\author{Ningyi Liao}
\authornotemark[1]
\email{liao0090@e.ntu.edu.sg}
\affiliation{%
  \institution{Nanyang Technological University}
  \country{Singapore}
}

\author{Siqiang Luo}
\authornote{Corresponding Author}
\affiliation{%
  \institution{Nanyang Technological University}
    \country{Singapore}
}
\email{siqiang.luo@ntu.edu.sg}

\author{Junfeng Liu}
\affiliation{%
  \institution{Nanyang Technological University}
  \country{Singapore}
}
\email{junfeng001@e.ntu.edu.sg}


\begin{abstract} 
Retrieval-Augmented Generation (RAG) enhances large language model (LLM) {generation quality} by incorporating relevant external knowledge. However, deploying RAG on consumer-grade platforms is challenging due to limited memory and the increasing scale of both models and knowledge bases. 
{In this work, we introduce {\PP}, a resource-efficient, self-adaptive RAG serving system integrated with LLMs, specifically designed for resource-constrained platforms.{\PP} exploits the insight that RAG retrieval and LLM generation impose different computational and memory demands, which in a traditional serial workflow result in substantial idle times and poor resource utilization. Based on this insight, {\PP} decouples retrieval and generation into parallel pipelines, incorporating joint memory placement and dynamic batch scheduling strategies to optimize resource usage across diverse hardware devices and workloads. Extensive experiments demonstrate that {\PP} adapts effectively to various hardware configurations and LLM scales,  achieving up to 3.6$\times$ speedup in average latency compared to serial RAG systems based on vLLM.}
\end{abstract}

\settopmatter{printfolios=true}
\maketitle
\let\clearpage\relax

\setlength{\textfloatsep}{6pt plus 0.8pt minus 2.6pt}
\setlength{\dbltextfloatsep}{6pt plus 0.8pt minus 2.6pt}
\setlength{\floatsep}{4pt plus 0.8pt minus 2.2pt}
\setlength{\intextsep}{4pt plus 0.8pt minus 2.8pt}
\setlength{\abovedisplayskip}{4pt plus 0.2pt minus 2.0pt}
\setlength{\abovedisplayshortskip}{4pt plus 0.2pt minus 2.0pt}
\setlength{\belowdisplayskip}{4pt plus 0.2pt minus 2.0pt}
\setlength{\belowdisplayshortskip}{4.0pt plus 0.2pt minus 2.0pt}

\section{Introduction}
The emergence of large language models (LLMs)~\cite{brown2020language, achiam2023gpt, touvron2023llama} has transformed diverse aspects of people's daily lives and production, fostering applications including programming assistants~\cite{chen2021evaluating, Github2022} and universal chatbots~\cite{Google2023, OpenAI2022}.
To ensure the quality and relevance of LLM responses in context-rich scenarios, Retrieval Augmented Generation (RAG)~\cite{asai2023retrieval} has become popular for invoking external knowledge bases, such as local and private data in companies~\cite{chen2024benchmarking,luo2023augmented,bayarriplanas2025paretooptimizedopensourcellmshealthcare, morris2023text}, into LLM inference. Local RAG utilizations, typically serving a small number of concurrent queries, focus on responding with low overall latency across these batches of data~\cite{fan2025minirag, prabhune2024deploying, wang2024speculative}.

In general, a RAG workflow first \textit{retrieves} data from the knowledge database according to user requests, then \textit{generates} responses by the LLM based on the extracted information. The integration of retrieval-based techniques, however, poses new challenges for comprehensive performance enhancements, especially as RAG systems are commonly deployed in environments with limited computational resources. 
This issue stems from the inherent tension between retrieval and generation components, which exhibit different characteristics in terms of computation needs and memory usage patterns. For example, while the retrieval operation benefits from higher database throughput through batch processing with larger batch sizes~\cite{Milvus,Chroma}, the generation operation usually demands a limited batch size to ensure low latency considering LLM tensor placement~\cite{song2024powerinfer,kwon2023efficient,sheng2023flexgen}. 



\begin{figure}[!t]
    \centering
    \includegraphics[width=0.95\columnwidth]{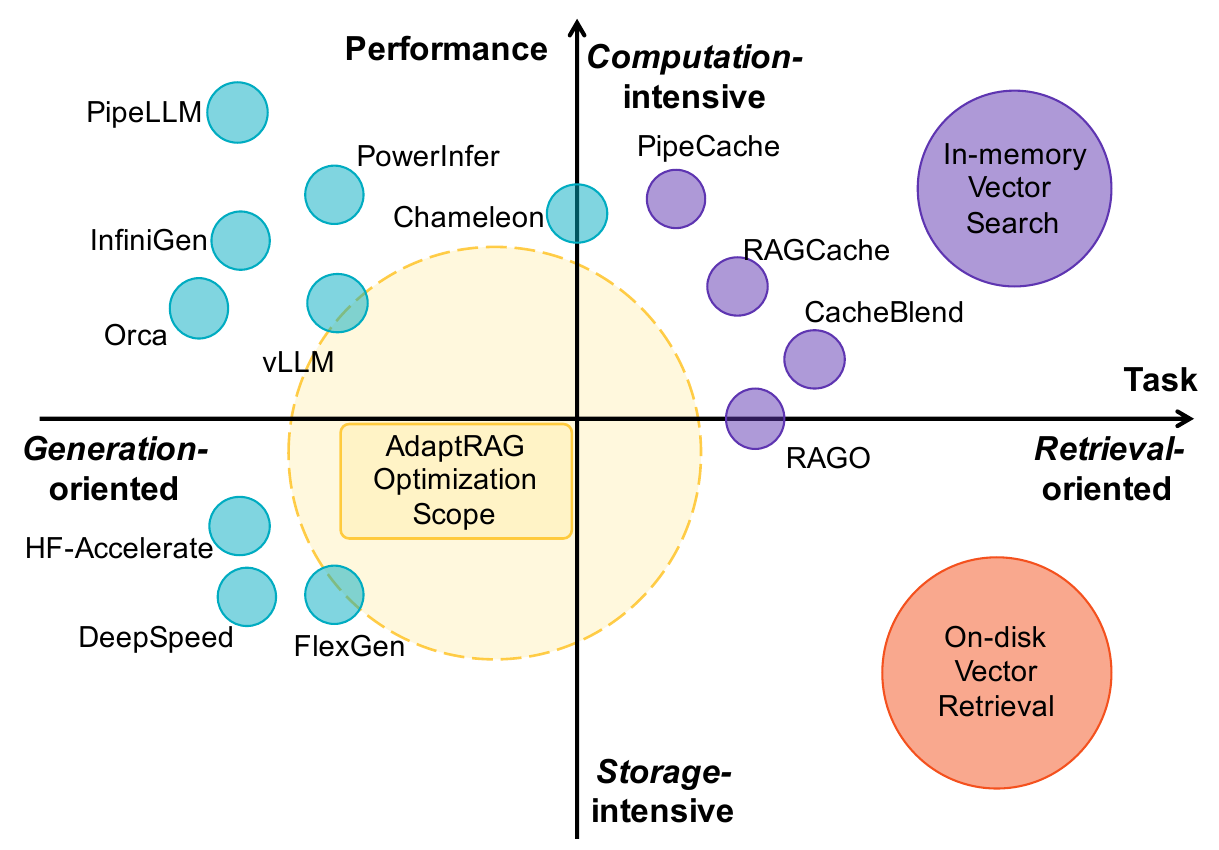}
    \caption{Representative techniques related to RAG serving, depicted by corresponding tasks and design objectives.}
    \label{fig:quadrant}
\end{figure}

In Figure~\ref{fig:quadrant}, we present existing studies representing LLM serving, RAG computation, and database retrieval optimization techniques from two dimensions: their intended \textit{tasks} focusing on retrieval or generation processes, and the \textit{performance bottleneck} in relation to computation and storage resources. Since most of these techniques focus only on a single component or design objective within the RAG system, directly applying them for RAG deployment on resource-constrained platforms leads to issues from two perspectives.

First, the RAG workflow features various data placements \textit{across devices}, especially concerning the GPU-CPU-disk storage hierarchy~\cite{sheng2023flexgen}. LLM serving frameworks usually leverage offloading techniques~\cite{aminabadi2022deepspeed,HuggingFace,kwon2023efficient,song2024powerinfer} in consumer-grade environments, transferring idle data to RAM and disk to save GPU memory. Vector databases used for knowledge retrieval also actively maintain partial or full-scale data in memory~\cite{douze2024faiss,aumuller2020ann,malkov2018efficient} or on disk~\cite{wang2021milvus,Chroma,jayaram2019diskann}. While both components tend to allocate more main memory for better performance, the available memory capacity is relatively limited compared to the sizes of the LLM and database. Excessive overhead risks memory thrashing or even exhaustion, resulting in impeded response latency.

Second, the processing and transmission of data \textit{over time} need to be carefully designed. The RAG pipeline manipulates a batch of data to be first computed from knowledge embeddings \cite{lewis2020retrieval,gao2023retrieval} by the CPU during retrieval, and then transferred to the GPU for generation. Additionally, the offloading strategy periodically relocates LLM tensors between the GPU and CPU. These operations exhibit intricate dependencies. Current studies on RAG serving enhancements \cite{jin2024ragcache,yao2024cacheblend} predominantly employ a straightforward serial processing strategy, which relies on the smooth execution of every phase. However, our evaluation reveals that overall latency is undermined by insufficient data utilization when computation and memory resources are constrained.
Hence, a comprehensive pipeline is necessary to minimize redundant data processing, waiting, and transmission time, while ensuring concurrent utilization of various devices.

In this paper, we propose {\PP}, a system for RAG deployment that integrates LLM and database components on a single consumer-grade GPU. As illustrated in Figure~\ref{fig:quadrant}, {\PP} encompasses a wide design space for collaboratively enhancing both computation and storage utilization, exploiting the benefits of different serving models adaptively toward lower latency. Specifically, {\PP} highlights three key designs for addressing the above challenges.

To effectively utilize storage across devices, {\PP} introduces an innovative \textit{joint memory placement} scheme integrating both retrieval and generation components. The system strategically optimizes placement across the entire storage hierarchy involving GPU, CPU, and disk, accounting for their distinct capacities and access speeds. This comprehensive maneuver improves device usage, achieving  $2\times$ improvement in average CPU utilization in practice compared to conventional serial RAG implementations. The dominant memory overhead elements, including database partitions and LLM tensors, can be coordinated through adjusting pivotal configuration parameters in \PP{}.

To enhance data utilization, \PP{} proposes a novel \textit{multi-pipeline RAG integration} that decouples CPU-bound retrieval and GPU-bound generation stages. This parallel execution on different computational devices effectively reduces the idle time from 80\% to 30\% of the overall latency in the RAG workflow. Furthermore, the fine-grained data flow manipulation loosens 
the component interdependencies, enabling a broader range of pipeline configurations.

Lastly, the joint and pipelined designs bring about the \textit{adaptive scheduling} ability of \PP{} for serving scenarios where static schemes fall short. Thanks to the dedicated data placement and pipeline execution modules, the impact on system workload can be effectively adjusted through a few performance-related parameters, especially batch size and resource allocation. The configurations in \PP{} serving are designed to be dynamically accommodated to the status of incoming queries and device utilization. To this end, \PP{} can adapt to different serving requirements by selecting the most suitable scheme dynamically, achieving peak latency reductions of up to 80\%.

We implement a \PP{} prototype, integrating the advanced FlexGen~\cite{sheng2023flexgen} as the offloading-based LLM serving system and Milvus~\cite{Milvus} as the high-performance vector database. We extensively evaluate its performance against popular RAG systems \cite{lewis2020retrieval,gao2023retrieval} deployed with LLM frameworks \cite{HuggingFace,kwon2023efficient} for consumer-grade platforms. Specifically, \PP{} is able to serve 8B-70B models under dynamic workload conditions, using only a single GPU with 12-24GB of memory and 176-256GB of main memory. Experimental results demonstrate that \PP{} reduces waiting and generation times by up to 20$\times$ and 5$\times$, respectively, achieving up to 3.6$\times$ and 11.7$\times$ boosts compared to baselines in overall average latency.

Our contributions can be summarized as follows:
\begin{itemize}[leftmargin=*]
    \item We propose \PP{}, a novel RAG system tailored for on-demand serving under memory constraints with a comprehensive design comprising computation and storage management for database retrieval and LLM generation.
    \item We introduce multi-pipelining for on-demand RAG serving. The novel LLM prefetching enables continuous enqueuing and dynamic data placement.
    \item We utilize adaptive configuration with online optimization, which comprehensively balances retrieval and generation overhead, as well as batch-wise latency.
    \item We implement \PP{}, which outperforms leading RAG deployments in resource-constrained settings by up to 3.6$\times$ in average latency.
\end{itemize}

\begin{figure}[t]
\centering
\includegraphics[width=1.0\linewidth]{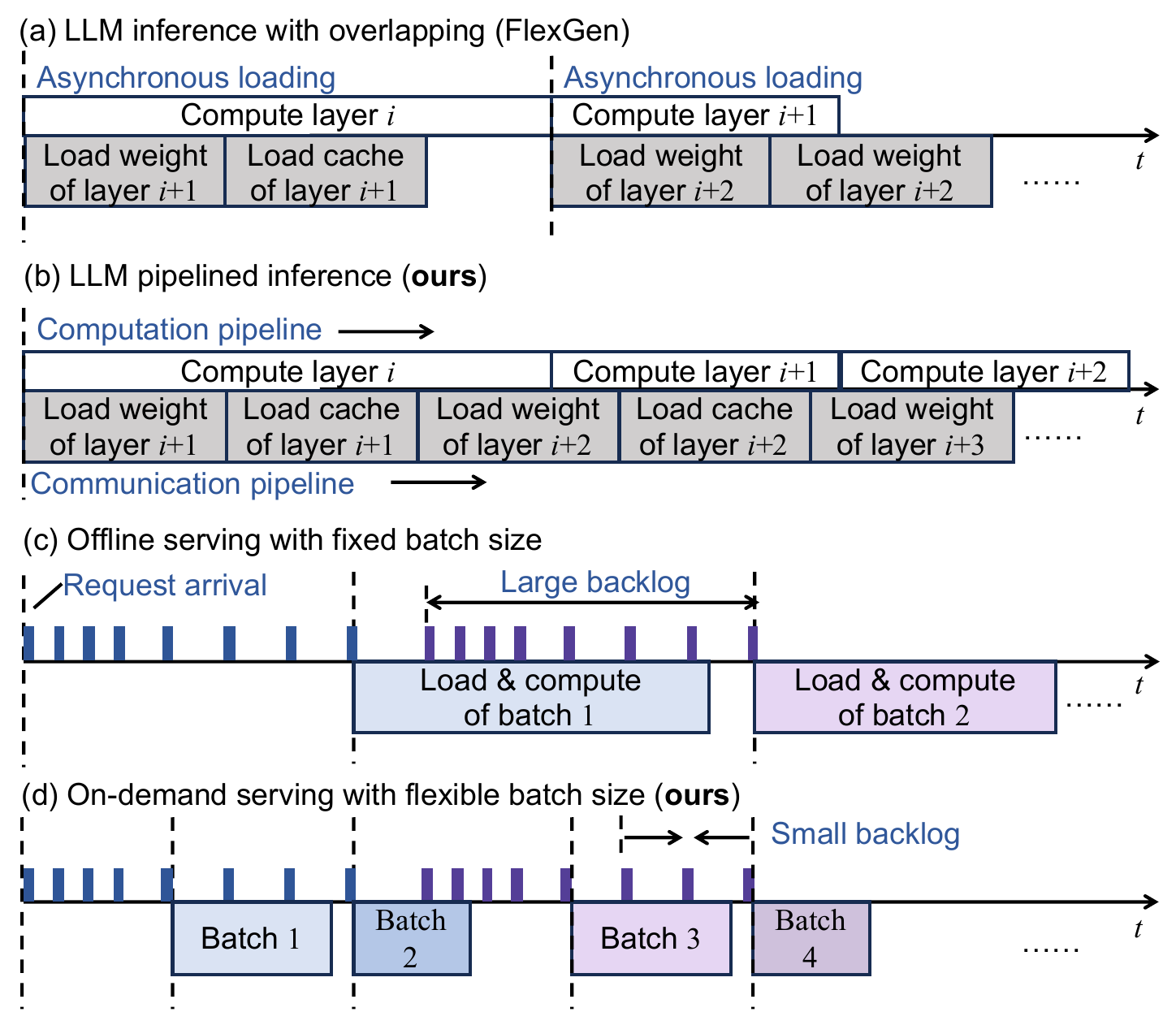}
  \caption{Pipelines in memory-intense RAG systems: \textbf{(a)} Standard overlapping LLM inference may misalign computation and prefetching due to CPU scheduling and compute jitter. \textbf{(b)} Our LLM pipeline separates computation and communication for continuous prefetching. {\textbf{(c)} Fixed batch scheduling accumulates larger backlogs under memory-intense conditions.  \textbf{(d)} Our backlog-aware batch scheduling adjusts flexibly to minimize backlogs. }} 
\label{fig:prefetching}
\end{figure}

\section{Background}


\subsection{Offloading-based LLM Inference} 
Typical LLM inference is performed in an auto-regressive manner~\cite{touvron2023llama,achiam2023gpt}. Given a user input, the model generates one token at a time by sequentially processing the data through a series of transformer blocks, namely \textit{layers}. 
Each layer comprises a self-attention mechanism and a feed-forward network with learnable parameters, namely \textit{weights}, to capture dependencies among tokens. The layer also manipulates a \textit{KV cache}, which stores previously computed key-value results in the attention module to reduce redundant computation throughout generation. As the data propagate through these layers, the model accumulates contextual information necessary for accurate token prediction. 

LLMs have seen a rapid growth in size in recent advancements, often exceeding the memory capacity of a single GPU~\cite{song2024powerinfer}. To alleviate the device requirement and enable LLM inference in consumer-grade platforms, offloading techniques~\cite{aminabadi2022deepspeed,HuggingFace,sheng2023flexgen,kwon2023efficient,song2024powerinfer} excel at transferring portions of the LLM tensors, including weights and KV caches, to RAM and disk, thereby bounding the GPU memory footprint.
Figure~\ref{fig:prefetching}(a) depicts the predominant practical LLM serving timeline~\cite{sheng2023flexgen,raistrick2024infinigen}. Upon computing the layer $i$, the system \textit{prefetches} the tensor of the next layer $i+1$ asynchronously. Communication time between GPU and host can be partially overlapped with computation to reduce inference latency.

However, in RAG applications, the memory occupation and transfer bandwidth are also vital for the retrieval performance. Irregular placement of model weights and cache may result in memory thrashing or congestion, which compels the system to resort to the slower disk storage and hinders the overall performance.

\subsection{On-demand RAG Serving} 
\label{sec:back/rag}
RAG presents a promising solution for supplying vast information from external knowledge bases to LLM inference, effectively expanding its contextual understanding and generation quality without retraining LLM. The retrieval ability is built on vector databases, where knowledge such as documents is processed into data chunks and embedded into vectors. During retrieval, these vectors are matched with the embedding of the given query through similarity search, and corresponding chunks with high relevance are used for subsequent LLM generation.

Contemporary knowledge bases are advantageous for managing massive volumes of data~\cite{joshi2017triviaqa,kwiatkowski2019natural} with embedding vectors spanning hundreds of gigabytes. This necessitates the on-disk storage scheme~\cite{wang2021milvus,Chroma}, where the database stores the majority of data chunks on disk, and only \textit{caches} relevant partitions to main memory for retrieval on demand. Although such a hierarchical storage layout allows for handling a larger scale of information within limited memory, the data transmission becomes another latency bottleneck, which is decisive for all succeeding RAG computations.

Canonical RAG frameworks \cite{LlamaIndex,LangChain} often encompass the \textit{offline} serving scenario, which handles incoming retrieval requests successively to guarantee individual end-to-end latency. However, due to vast knowledge base size and auto-regressive nature of LLMs, processing times of both retrieval and generation are not bounded. In \textit{online} circumstances~\cite{kwon2023efficient,jin2024ragcache,yao2024cacheblend}, where concurrent queries occur, requests must wait for the preceding ones to release resources. Consequently, the overall latency among all queries is not satisfactory.

Recent works improving RAG serving~\cite{jin2024ragcache,yao2024cacheblend,zhu2024accelerating,jiang2024piperag,jimenez2024hipporag} mostly focus on the computation-intensive components in the system, assuming abundant memory resources and efficient online serving. The above temporal and device-wise challenges are highly underexplored in these works. We therefore regard them as orthogonal to this study.

\subsection{Batch Scheduling} 



{\color{black} Batching strategies play a crucial role in optimizing the performance tradeoff between throughput and latency in both LLM generation and vector retrieval in RAG systems. In LLM serving architectures, these strategies typically fall into two categories: online and offline processing. Online systems like vLLM~\cite{kwon2023efficient} implement iteration-level scheduling by allocating a predefined maximum batch size and dynamically adjusting batch composition to minimize average latency. Conversely, offline systems like FlexGen~\cite{sheng2023flexgen} prioritize throughput by employing larger batch sizes.  }

{\color{black} Modern vector databases~\cite{Milvus,Chroma,douze2024faiss} support efficient batch processing of multiple embedding queries, allowing them to handle high query volumes while maintaining low latency. For example, Milvus allows sending a batch of vectors in one API call instead of processing them individually. This minimizes network round-trips and leverages bulk processing optimizations in the database engine.  }

{\color{black} Existing RAG serving frameworks~\cite{jin2024ragcache,yao2024cacheblend,ram2023context} leverage LLM batching strategies to optimize overall performance. Systems like RAGCache~\cite{jin2024ragcache} and CacheBlend~\cite{yao2024cacheblend} build upon vLLM's~\cite{kwon2023efficient} batch scheduling mechanisms to improve latency. RAGCache implements a maximum batch size approach to pursue lower average latency, while CacheBlend evaluates performance across different batch scheduling without specifying batch sizes for on-demand scenarios. These approaches demonstrate that effective batching techniques are crucial for achieving low latency in RAG systems.}

\section{Motivation}

\subsection{Problem Formulation}




In this study, we aim for the problem of resource-constrained RAG serving, which comprises three principal components: a vector database with capacity $V$, an LLM of size $M$, and a running cache of size $C$. 
Function $\mathcal{P}_{\theta}(V, M, C)$ returns a $d$-dimensional vector, representing the allocated memory for the system \textit{components} to $d$ storage devices, which is controlled by the memory placement strategy $\theta$. The LLM execution requires additional storage for hosting the \textit{data}, such as KV cache, which is characterized by function $\mathcal{W}(B, M)$, where $B$ is the generation batch size. The device storage capacity is bounded by vector $D \in \mathbb{R}^d$.

The online serving scenario features incoming query rate as a variable process $\lambda(t)$ through time $t$. 
Our objective is to minimize the expected latency $\mathcal{L}(B, \theta, \lambda(t))$ through joint optimization of batch size $B$ and placement strategy $\theta$ under the condition of memory bound of each device. This yields the following constrained optimization problem:
\begin{equation}
\begin{split}
\min_{B, \theta} \quad & \mathcal{L}(B, \theta, \lambda(t)), \\
\text{s.t.} \quad & \mathcal{P}_{\theta}(V, M, C) + \mathcal{W}(B, M) \preceq D,
\end{split}
\end{equation}
where the partial ordering $\preceq$ indicates component-wise inequality, indicating that the memory overhead on each device, e.g., VRAM, RAM, and disk, remains within its respective capacity constraint.

\begin{figure}[t]
\centering
\includegraphics[width=1.0\linewidth]{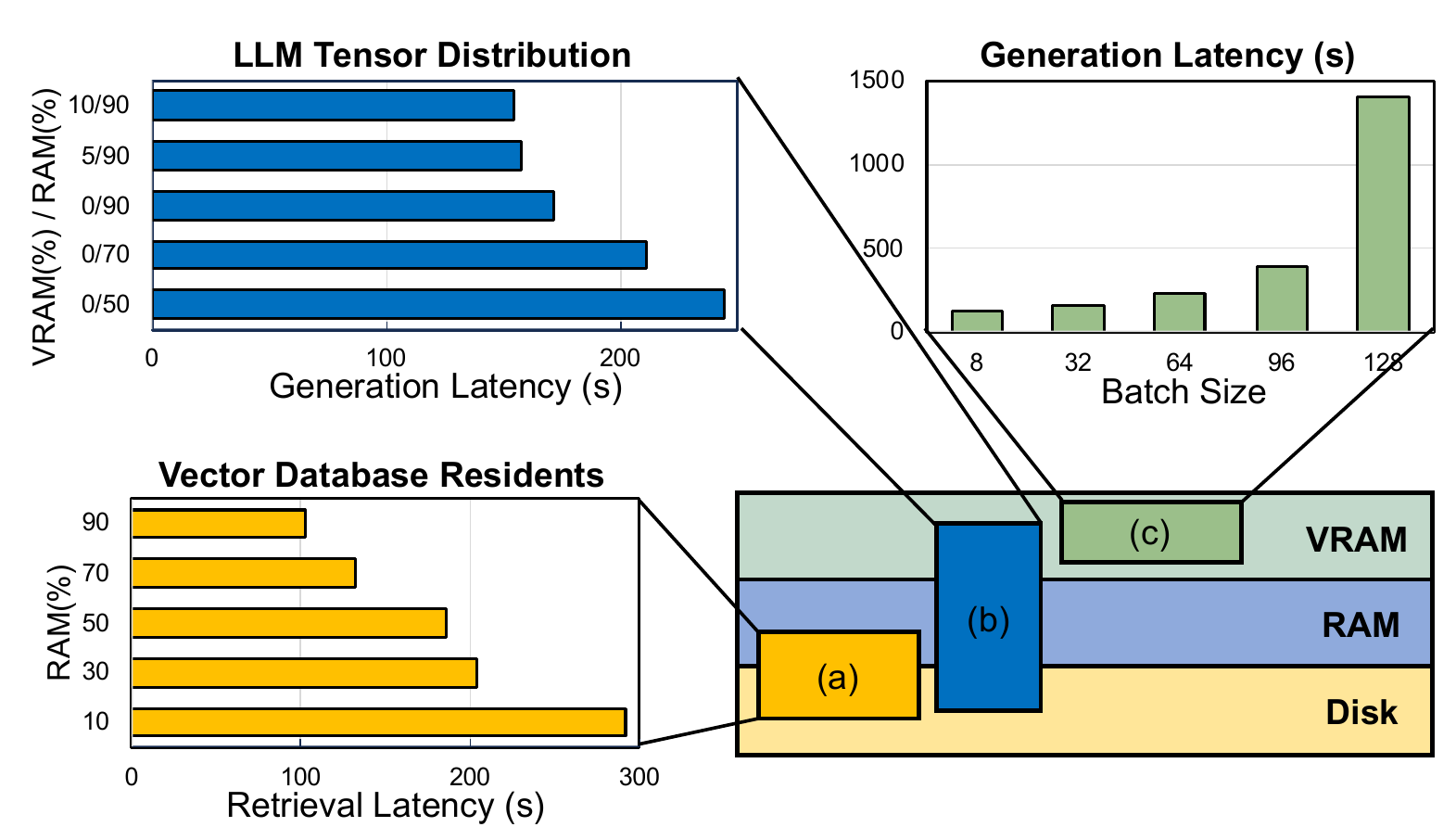}
  \caption{Dissecting an online offloading-based RAG system. \textbf{(a)} LLM tensor placement. \textbf{(b)} vector database residents. \textbf{(c)} compute workspace scheduling. }
\label{fig:dissect}
\end{figure}

\subsection{Challenges in RAG Systems}
\label{sec:challenge}
Prevailing RAG serving frameworks \cite{jin2024ragcache,yao2024cacheblend,jiang2024piperag} primarily enhance computation and memory placement on GPU while assuming abundant RAM and small-scale database retrieval. Regarding the LLM component, offloading-based strategies are directly employed to tackle the device constraints and boost computation. 
By deploying representative RAG systems on consumer-grade platforms, we observe that the serving performance exhibits a more intricate landscape, rendering new challenges for optimizing latency.

\noindentparagraph{C1: Significant impact from memory-intensive operations.} 
{\color{black} While vLLM~\cite{kwon2023efficient} facilitates KV caches placement across devices, systems like RAGCache~\cite{jin2024ragcache} and CacheBlend~\cite{yao2024cacheblend} still require substantial computational resources—two H800 GPUs (2$\times$80GB VRAM) and two A40 GPUs (2$\times$48GB VRAM) respectively—to employ the LLaMA2-70B model as their LLM component. The memory requirements intensify further when handling large-scale knowledge bases that necessitate on-disk vector databases. Consequently, adapting these existing paradigms to common single-GPU environments remains non-trivial. We systematically evaluate this deployment challenge by varying three pivotal factors affecting serving latency: the number of cached partitions in the vector database, the proportion of LLM tensors placed on the GPU, and the generation batch size.}

Figure~\ref{fig:dissect} reveals that the respective and overall latency are dominated by configurations on memory placement. Caching more database partitions in RAM from disk can favorably reduce retrieval latency, while an unaggressive offloading strategy for keeping more model tensors on GPU benefits generation efficiency. 
Notably, the large generation batch size $B=128$ significantly escalate LLM serving time. This is caused by the increased footprint of KV cache, forcing LLM tensors and vector database partitions to be offloaded to lower memory hierarchies. This observation indicates potential resource race between RAG components when they are not properly configured in a \textit{comprehensive} manner.

\begin{figure}[t]
\centering
\includegraphics[width=1.0\linewidth]{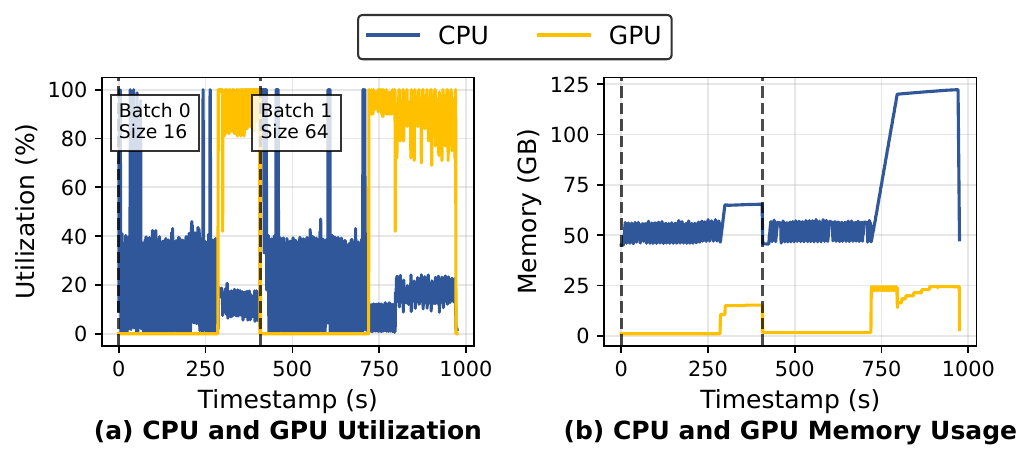}
  \caption{CPU and GPU utilization and memory usage vary in a serial retrieval and generation mode when using different batch sizes under a static memory allocation policy.}
\label{fig:profiling}
\end{figure}

\noindentparagraph{C2: Extra waiting time beyond LLM generation.} 
Despite the memory-side overhead, RAG processing essentially demand a considerable running time. Figure~\ref{fig:profiling}(a) assesses the device utilization for retrieval on CPU and generation on GPU. The average end-to-end latency within the first batch is approximately 400 seconds, while queries in the second batch reach up to 1000 seconds, which is largely unsatisfactory in practical applications. 

We mainly identify two patterns of idle times that increase the latency. The \textit{external} latency is relevant to the backlog phenomenon of batch processing when resources are constrained, in that some requests may already arrive during processing the first batch but have to wait for the previous completion and only be processed in the next batch. This issue affects the end-to-end latency of certain queries and causes long waiting times in extreme cases. 

Meanwhile, the \textit{internal} latency roots in the underutilization of GPU and CPU within each batch. Retrieval operations primarily utilize CPU for vector matching in the database, while generation predominantly leverages GPU compute capacity. Hence, without \textit{pipelining}, these devices are left with long idle times waiting for the computation on the other one, undermining system throughput and efficiency.

\noindentparagraph{C3: Irregular workload of on-demand serving.} 

{\color{black} Unlike offline RAG serving, where system configurations such as batching policies can be predetermined, on-demand use cases must accommodate variable request arrival rates~\cite{wang2024burstgpt,patel2024splitwise,stojkovic2024dynamollm}. Limited computational and memory resources further complicate on-demand scheduling. Specifically, fluctuating request patterns during peak and off-peak periods form batches of varying sizes, which entail distinct computational and memory utilization patterns~\cite{pang2025optimizing,patke2024one}. As illustrated in Figure~\ref{fig:profiling} (b), variations in batch size significantly affect both GPU and CPU memory overhead. For instance, CPU memory utilization increases from approximately 60GB to 120GB during the generation phase. Moreover, retrieval and generation exhibit different characteristics as batch size increases: retrieval time remains relatively constant even at batch size 64, while generation time doubles.
}

{\color{black} This insight suggests the need for an \textit{adaptive} serving mode that configures memory allocation and batch scheduling based on online workload conditions. However, conventional online batch schedulers such as vLLM do not adequately address the requirements of memory-intensive RAG systems. This mismatch occurs because: (1) retrieval and generation demonstrate different scaling behaviors as batch size increases, making uniform batching across both processes suboptimal; (2) in memory-intensive scenarios where numerous requests are already backlogged before each batch, fine-grained request replacement policies may paradoxically degrade performance~\cite{kwon2023efficient}. Consequently, we need to carefully model the costs of both retrieval and generation processes, and design backlog-aware batch scheduling strategies.}

\section{{\PP}}
In this section, we present {\PP}, an adaptive multi-pipeline RAG system for large-scale data and constrained GPU and memory devices. We first outline the new design space offered by \PP{} for organizing the RAG workflow comprehensively (Section~\ref{sec:overview}). Then, we respectively introduce the novel components and their integration, including the memory placement over devices (Section~\ref{sec:placement}) and the offloading-based inference systems (Section~\ref{sec:prefetching}). Lastly, we showcase the adaptive tuning of the configuration space through a two-step strategy (Section~\ref{sec:tuning}).

\subsection{Design Space Overview}
\label{sec:overview}
With the objective to deploy RAG systems in consumer-grade environments with low latency, we attempt to offer a new schema for jointly designing the RAG components, their collaboration, and dynamic configuration during serving. This novel design space not only enhances \PP{} through more efficient use of resources, but also contributes to the broader community by exploring how such designs can improve both the effectiveness of retrieval and generation, as well as the balance between computational and storage demands in RAG systems.

\begin{figure}[t]
\centering
\includegraphics[width=1.0\linewidth]{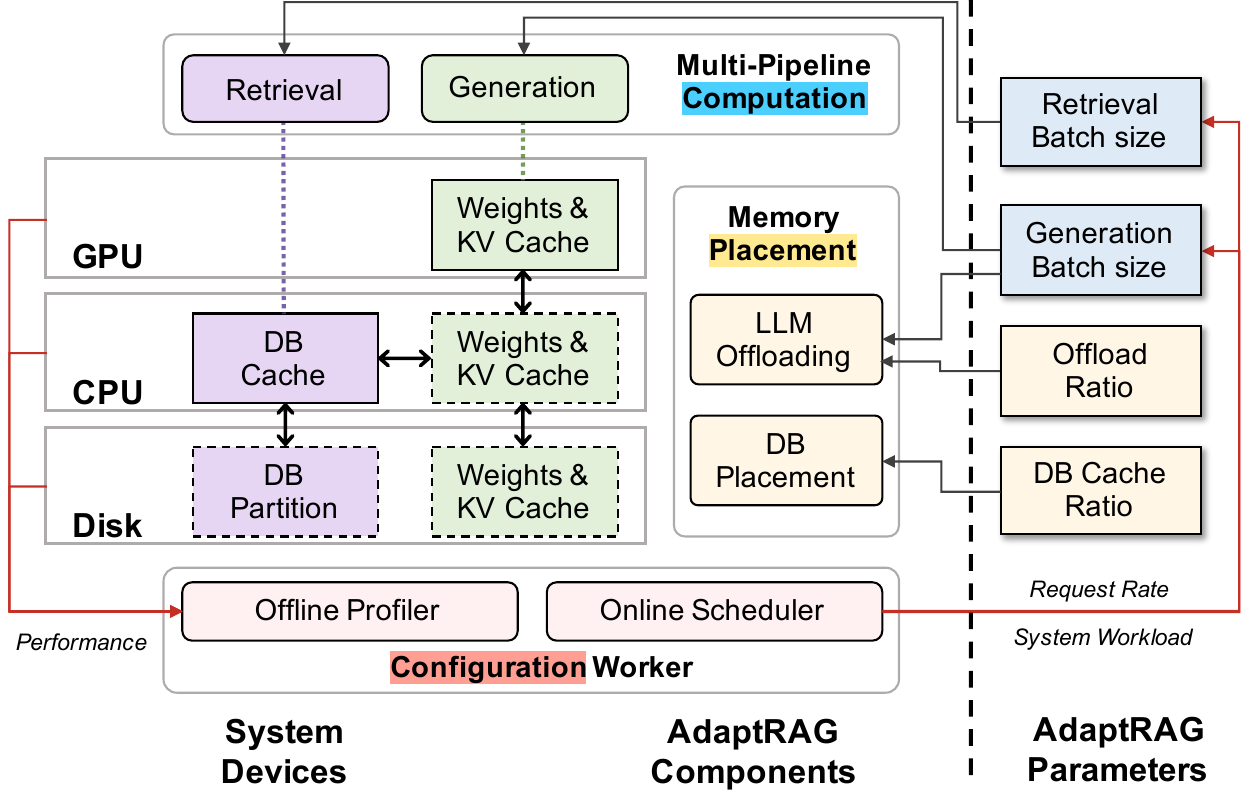}
  \caption{Overview of \PP{}.}
\label{fig:design}
\end{figure}

\PP{}'s novel designs aim to address the critical challenges in Section~\ref{sec:challenge}. As illustrated in Figure~\ref{fig:design}, modules in \PP{} are tunable during processing, allowing for accommodation to various serving conditions and resources through adjusting pivotal parameters. To showcase the overall design space, we respectively introduce the three key components and their configurable factors in \PP{} as follows, while detailed designs are elaborated in later subsections. 

\noindentparagraph{Joint pipeline: Hierarchical memory placement.}
\PP{} features a novel pipelined scheme, which operates retrieval and generation processes as separate workers. Incoming user requests are first placed into the retrieval queue, which is actively formed as batches for database query. The resulting contexts containing user requests and retrieved document chunks are transferred to the generation queue, where LLM batch processing is performed to generate response texts. {Due to the irregular completion time of each batch, the retrieval and generation batches can be formed with different requests, which loosen the dependency of processing orders among them.}

The pipelining approach is advantageous for explicit management across different hardware resources. As for \textit{computation}, retrieval operations primarily utilize CPU resources, while generation tasks leverage GPU capabilities. Decoupling these two pipelines allows for parallel execution with higher device utilization. Regarding \textit{storage}, partitions of the vector database, LLM parameters, and KV caches can be dynamically loaded or offloaded across GPU, CPU, and disk memory hierarchies to ensure efficient computation under capacity constraints. 

\noindentparagraph{Offloading prefetch: Dynamic batch size.}
To specifically deploy RAG with a single GPU, LLM offloading is necessary for dynamically scheduling model weights and KV caches in runtime. \PP{} implements an innovative prefetching mechanism in offloading-based LLM serving, which is tailored for the pipeline RAG system. To enhance generation efficiency, we simultaneously conduct computation and communication operations to reduce idle time. 

{\textit{Processing batch size} is a critical parameter to balance generation throughput and latency. Our configuration intends to tune the ideal batch size for achieving low latency under varying request arrival rates.} Our system further adjusts batch scheduling based on workload patterns in real time to efficiently address processing backlogs. Date placements within the memory hierarchy are adjusted accordingly to accommodate the batch scheduling between generation phases.

\noindentparagraph{Adaptive configuration: Workload feedback.}
By integrating the tunable components, a configuration layer is applied to adapt the system to compute workload throughout serving. The configuration in \PP{} is specifically tuned by two aspects of factors. An active profiler first iteratively explores the \textit{configuration space} to identify optimal settings that minimize average latency while balancing the dual pipelines a priori. During the operational phase, a batch scheduler is designed for selecting the optimal batch size based on the actual \textit{backlog}. Based on the configuration, the system schedules the other components accordingly to reduce average latency across diverse workloads.

\subsection{Hierarchical Memory Placement}
\label{sec:placement}
\begin{figure}[t]
\centering
\includegraphics[width=1.0\linewidth]{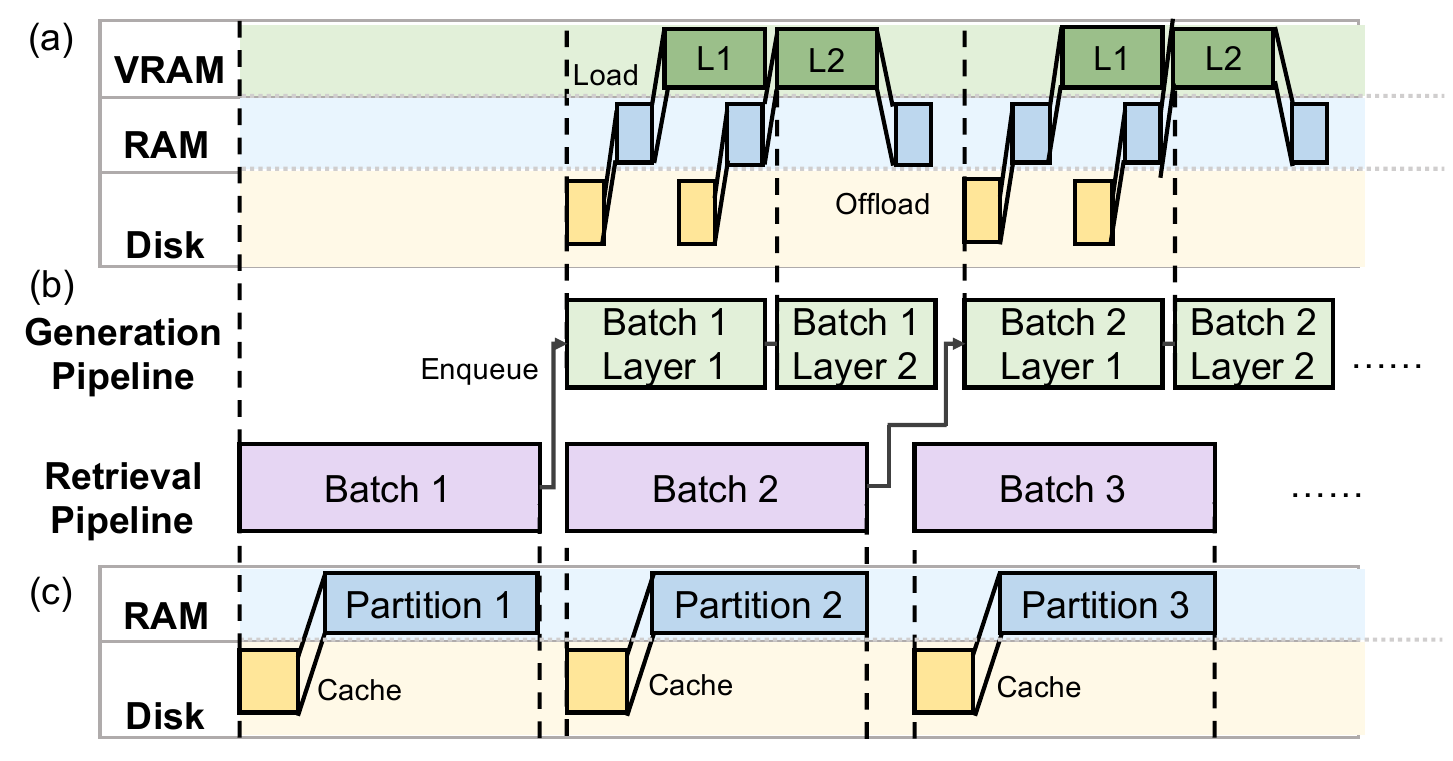}
  \caption{Timelines of computation and memory operations in \PP{}: \textbf{(a)} memory placement for LLM weights and KV cache; \textbf{(b)} computation workspace of multi-pipeline; \textbf{(c)} memory placement for database partitions.}
\label{fig:memory}
\end{figure}

Conventional RAG workflows exhibit strict dependencies between retrieval and generation phases, creating inefficiencies in data processing and storage.
{\color{black} Though pipelining retrieval and generation is not new, we note that implementing effective pipelining under memory-intensive scenarios still
presents challenges. {\PP} addresses these challenges by comprehensively managing memory allocation across the complete memory hierarchy to prevent resource imbalances and hardware underutilization. This approach optimizes several critical components as shown in Figure~\ref{fig:memory}: }


\noindentparagraph{Database partitions.}
{\color{black} Our system implements sophisticated partition management for vector retrieval, allowing partitions to be dynamically loaded or released according to current operational demands. When the system determines that adjusting the number of resident partitions would optimize performance, it strategically identifies candidates for release and acquisition, executing these operations between consecutive batches of retrieval tasks.}

\noindentparagraph{LLM weights and KV cache.}
{\color{black} For both model weights and KV cache management, we employ a percentage based memory placement strategy that dynamically allocates resources across the memory hierarchy. For instance, if analysis determines an optimal model weight distribution of 50\% GPU, 50\% CPU, and 0\% disk utilization, the system systematically relocates tensors according to this new ratio for the subsequent generation batch. This distribution is governed by configuration that specify the placement of each tensor.}

\noindentparagraph{Computation workspace.}
{\color{black} We implement adaptive computation workspace allocation based on batch size requirements. As batch composition changes, the memory necessary for intermediate computations during model inference must be correspondingly adjusted, including attention matrices and hidden states. The system calculates memory requirements based on the current batch configuration and allocates resources accordingly before processing each generation batch.}

Between operational batches, our system employs a sophisticated lock mechanism to ensure that memory reorganization proceeds safely and efficiently. The retrieval and generation workers operate with independent locking protocols, enabling concurrent operation while preventing resource conflicts during configuration updates. Further details on tensor and partition transfer mechanisms can be found in Section~\ref{sec:implement}. This comprehensive approach allows the system to respond dynamically to fluctuating workloads without manual intervention, maximizing resource utilization while maintaining optimal performance characteristics.

 \subsection{LLM Prefetching Pipeline}
\label{sec:prefetching}
{\color{black} {\PP}'s prefetching module optimizes data movement between GPU, CPU, and disk during offloading-based LLM inference. While existing approaches employ fixed prefetching strategies (Figure~\ref{fig:prefetching}(a)), they suffer from inefficiencies in resource-constrained environments due to CPU scheduling and compute jitter~\cite{agrawal2024etalon,agrawal2024metron}.These variations create misalignment between computation and prefetching operations, resulting in resource underutilization. Our enhanced approach (Figure~\ref{fig:prefetching}(b)) builds upon and enhances this strategy.}

\noindentparagraph{Data flow over time.}
\PP{} implements an enhanced asynchronous tensor management system using a thread-based execution model. Rather than restricting prefetching to only the immediate next layer, we continuously enqueue subsequent layers into a prefetching queue—subject only to available memory resources (as illustrated in Figure~\ref{fig:prefetching}(b)). Dedicated thread pools and multiple CUDA streams manage these concurrent data transfers, ensuring that the required tensors are available for future computation iterations.

This approach effectively minimizes idle time by accounting for computation and loading variations while also dynamically adjusting prefetching strategies based on available memory resources in both prefill and decoding phases. This adjustment is necessary because prefill and decoding phases have fundamentally different memory usage patterns, as demonstrated by FlexGen~\cite{sheng2023flexgen}. During prefill, activation memory requirements are significantly higher since all input tokens must be processed simultaneously, leaving less room for prefetching queues. Conversely, during decoding, only a single new token is processed at each step, resulting in lower activation memory usage and allowing for more aggressive prefetching strategies. Our system leverages this insight to optimize queue capacity, using a more conservative prefetching distance during prefill and increasing prefetching aggressiveness during decoding. 

\noindentparagraph{Tensor placement across devices.}
The prefetching module also integrates closely with the hierarchical memory management described in Section~\ref{sec:placement}. It can load and offload tensors across the memory hierarchy and dynamically adjust batch sizes and memory allocations in response to changing workloads. This coordination between components enables {\PP} maintaining consistent performance even under challenging operational conditions.

\subsection{Adaptive Serving Configuration}
\label{sec:tuning}
To tune the design space, we propose a two-step paradigm. First offline, we employ targeted active profiling to systematically collect performance characteristics of hardware under complex parallel processing; Second online, we provide a backlog-aware batch scheduling approach which is designed to optimize the batch scheduling during the runtime phase.

\noindentparagraph{Active Profiling (Offline).} Typical RAG systems often adopt a static configuration, which fails to adapt to different workload characteristics and hardware constraints. To address this limitation, we propose an active profiling approach for preparation that dynamically configures the memory hierarchy. Our method iteratively explores the configuration space to find an optimal balance between the retrieval and generation pipelines. The core insight of our active profiling approach is that the optimal configuration should balance the latencies of both pipelines. For example, if retrieval time exceeds generation time, we adjust by increasing residential partitions on CPU and reducing LLM tensor allocations on CPU. This balancing act operates under hardware memory constraints that can be expressed as:
\begin{align}
w_{gpu} \cdot W_{total} + c_{gpu} \cdot C(B) + H(B) \leq M_{gpu} \\
w_{cpu} \cdot W_{total} + c_{cpu} \cdot C(B) + P \cdot M_p \leq M_{cpu}
\end{align}
\noindent where $w_{gpu}$ and $w_{cpu}$ represent the percentage of model weights stored in GPU and CPU memory respectively, $W_{total}$ is the total size of model weights, $c_{gpu}$ and $c_{cpu}$ represent the percentage of KV cache stored in GPU and CPU memory, $C(B)$ is the cache size for batch size $B$, $H(B)$ is the size of computation workspace,
$P$ is the number of resident partitions, $M_p$ is the size of each partition, and $M_{gpu}$ and $M_{cpu}$ represent the available GPU and CPU memory.

Instead of constructing a detailed analytical model, we sample configurations at strategic intervals. While ideally, we would explore all pairs of concurrent retrieval and generation batch sizes to minimize $\max(t_{retrieval}, t_{generation})$, this is computationally expensive. Recognizing that retrieval and generation exhibit different batching behaviors, particularly under memory pressure, we simplify the search. We observe that retrieval cost is dominated by partition loading (see Section~\ref{sec:challenge}). Therefore, for a fixed number of resident partitions within a given configuration, its value remains nearly constant across varying retrieval batch sizes. Leveraging this, we effectively treat $t_{retrieval}$ as a constant factor and focus the active profiling on tuning the generation batch size. This targeted exploration efficiently identifies the generation batch size that best balances the pipeline latencies, accounting for the complex memory hierarchy interactions more effectively than exhaustive search or purely analytical models.

\noindentparagraph{Backlog-aware Batch Scheduling (Online).} 
Resource-constrained RAG systems are particularly susceptible to overloading, which frequently results in request backlogs that existing scheduling approaches struggle to handle effectively.
To specifically address these  challenges, we analyze the correlation between batch size and average latency. Let $T(B)$ represent the processing time for a batch of size $B$:
\begin{equation}
T(B) = a \cdot B^c
\end{equation}
where $a>0$ and $c \geq 0$ are constants. For a single batch containing all $n$ requests, the average latency is:
\begin{equation}
L_{1} = T(n) - \frac{1}{n}\sum_{i=1}^{n}t_i
\end{equation}
When dividing requests into $k$ equal batches of size $\frac{n}{k}$, the average latency becomes:
\begin{equation}
L_{k} = \frac{k+1}{2} \cdot T\left(\frac{n}{k}\right) - \frac{1}{n}\sum_{i=1}^{n}t_i
\end{equation}
The maximum batch size is optimal when $L_{1} \leq L_{k}$, giving:
\begin{equation}
T(n) \leq \frac{k+1}{2} \cdot T\left(\frac{n}{k}\right) \Rightarrow 2 \cdot k^c \leq k+1
\end{equation}
\noindent For example, when $k=2$ (splitting into two equal batches):
\begin{equation}
2 \cdot 2^c \leq 3 \Rightarrow c \leq \log_2\left(\frac{3}{2}\right) \approx 0.585
\end{equation}

Therefore, using the maximum batch size minimizes average latency when the processing time scales sublinearly with batch size at a rate less than the threshold determined by $k$, while dividing into smaller batches is more efficient when processing time scales more rapidly with batch size. 

In practice, our system implements this insight by evaluating different batch sizes at runtime based on samples collected during the active profiling phase. For example, when there are 64 requests in the queue, we assess the latency for batch sizes of 8, 16, 32, and 64 using the above formulas to determine the best configuration. Note that retrieval and generation workers adopt different batch scheduling strategies due to their distinct characteristics. Retrieval tasks typically perform well with larger batches, as they primarily rely on the CPU for vector operations. In contrast, LLM inference offloads more tensors to lower memory hierarchies with larger batches, causing processing time to scale more rapidly with batch size. Our approach accounts for these differences by optimizing each pipeline independently based on its performance characteristics.

\begin{figure*}[t!]
\centering
\includegraphics[width=0.95\linewidth]{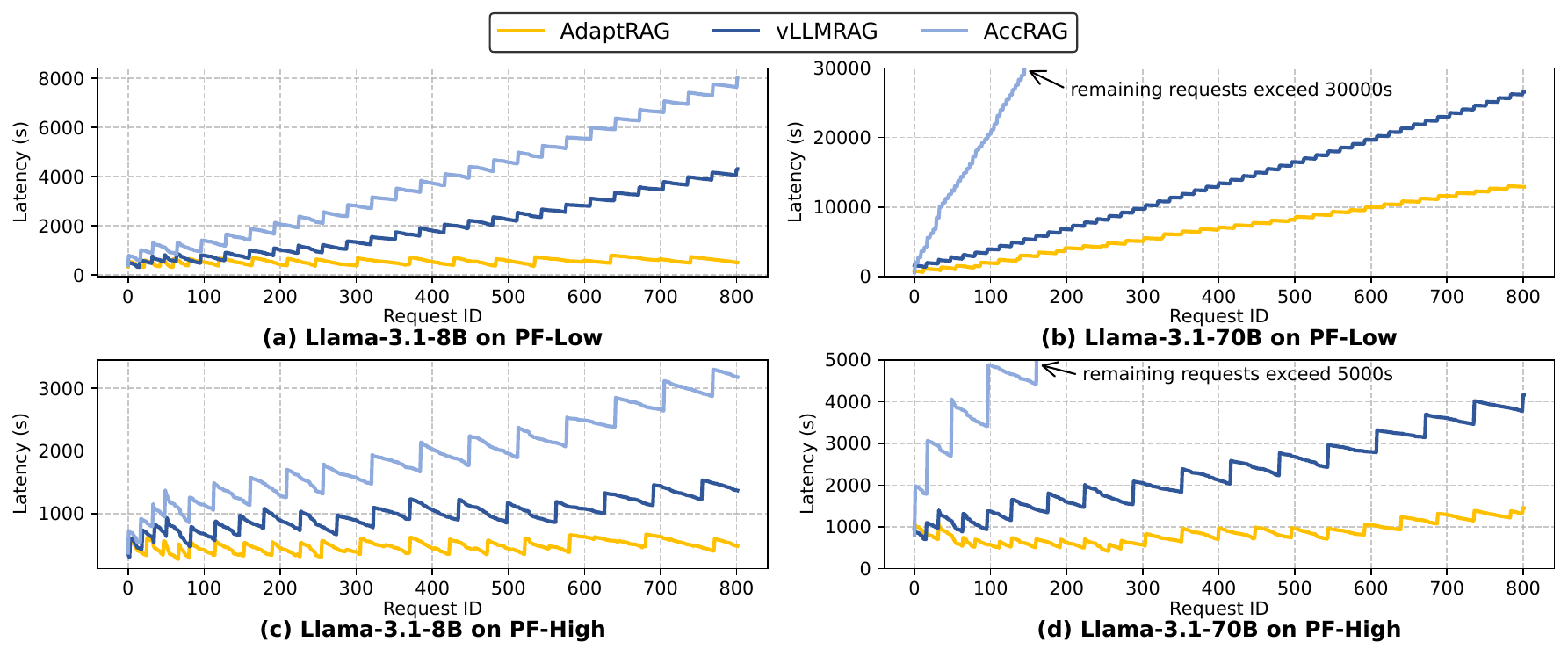}
\caption{Request latency under a dynamic workload. On the x-axis, the arrival rate increases from left to right: approximately 0–80 indicates 4 requests per minute, 80–240 corresponds to 8 requests per minute, 240–480 to 12 requests per minute, and 480–800 to 16 requests per minute. The linear segments between two distinct turning points represent a generation batch, since nearly uniform arrival times lead to similar completion times and thus a linear latency profile.}
\label{fig:perf}
\end{figure*}

\section{Implementation}
\label{sec:implement}
We have implemented a prototype of {\PP} in approximately 5000 lines of code. Our knowledge base is built upon the \texttt{Document} class in LangChain~\cite{LangChain}, a widely used framework for integrating LLMs into applications. For storage, we use Milvus~\cite{wang2021milvus}, a popular vector database recognized by the GitHub community~\cite{Milvus}, to manage document chunks and associated embeddings. Additionally, our LLM prefetching module is built on FlexGen~\cite{sheng2023flexgen}, which is specifically designed for offloading-based inference on a single GPU.

\noindentparagraph{Lazy dynamic transfer.} Given our adaptive computation workspace and dynamic memory hierarchy, it is essential to transfer tensors during LLM inference when policy changes occur. We adopt a lazy transfer approach: rather than transferring model weights between CPU and GPU memory synchronously, we perform these transfers on a background CUDA stream and only synchronize when a weight tensor is imminently required for computation. Moreover, once a weight tensor has been offloaded to disk, we retain its unique file name and reuse it across generations, ensuring that each weight tensor incurs an offload cost only once. Since our weight placement policy adjusts incrementally, only a small subset of tensors require transfer during policy updates—this mechanism applies to both profiling and runtime phases. For KV cache, we clear all cached data from the memory hierarchy and reinitialize it at the start of each generation, following the same pattern used for hidden states and attention masks. Similarly, for vector database partitions, we adopt a lazy strategy by releasing and loading partitions during the retrieval phase, thereby keeping the overall cost of policy transfer minimal.

\noindentparagraph{Fault tolerance.} Despite provisioning for additional memory on both the GPU and CPU, resource constraints may still lead to out-of-memory (OOM) errors. To mitigate this, we have incorporated fault-tolerant mechanisms in both our offline profiling and online runtime phases. We checkpoint intermediate retrieval chunks for each partition; should a retrieval failure occur, the system reverts to the most recent successful checkpoint. For LLM inference, a checkpoint is maintained at the beginning of each generation. In the event of a GPU or CPU OOM during generation, we first attempt to reinitialize the KV cache by shifting more tensors to a lower memory tier; if that is insufficient, we offload a portion of the weight tensors to lower memory. This design minimizes the need for a full restart and reduces the overhead associated with additional I/O operations.

\section{Evaluation}
In this section, we evaluate the performance of {\PP} compared to baseline systems under various experimental configurations, including different hardware setups, model sizes, and workloads.

\subsection{Experimental Setup}
\noindentparagraph{Hardware.}  The experiments are conducted on two platforms representing relatively high- and low-end scenarios:
\begin{itemize}[leftmargin=*]
    \item \textbf{PF-High:} Intel Xeon 4314 processor (63 cores at 2.4 GHz), 256 GB of host memory, an NVIDIA A30 GPU (24 GB), an 8 TB SSD, and PCIe 4.0 interface (64 GB/s bandwidth).
    \item \textbf{PF-Low:} Intel Xeon 6238R processor (111 cores at 2.2 GHz), 176 GB of host memory, an NVIDIA A5000 GPU (12 GB), a 2 TB SSD, and PCIe 3.0 interface (32 GB/s bandwidth).
\end{itemize}

\noindentparagraph{Dataset and Models.} For the knowledge base, we construct a vector database by chunking and embedding documents from the TriviaQA dataset~\cite{joshi2017triviaqa}, which was selected for its high document volume. TriviaQA is a reading comprehension dataset containing over 650,000 question and answer tuples. The vector database is partitioned into 32 segments, with each partition being approximately 8 GB in size, resulting in a total database size of approximately 256 GB. For the LLMs, we adopt two representative sizes: Llama 3.1 with 8 billion and 70 billion parameters~\cite{touvron2023llama}. We chose this range (from 8B to 70B) because these model sizes are not only popular among LLMs as evidenced by leaderboard~\cite{LMSYS,kwon2023efficient}, but also because they can be reasonably executed on a single consumer-grade GPU while allowing most baseline methods to function normally for fair comparison.

\noindentparagraph{Workloads.} Each request in our workload is synthesized from a question in the TriviaQA dataset. Under the default settings, the top 5 most relevant document chunks are retrieved from the vector database and concatenated with the original question to form the final input for the LLM. Since the dataset does not include timestamps, we simulate request arrival times using a Poisson distribution with varying request rates, following \cite{kwon2023efficient,yu2022orca}. The entire workload is divided into 20-minute intervals, with each interval assigned a specific request rate. By default, we use four intervals with request rates of 4, 8, 12, and 16 requests per minute, covering both sustainable and overload scenarios. 

\noindentparagraph{Baselines.}
As discussed in Section~\ref{sec:challenge}, most existing RAG optimization systems \cite{jin2024ragcache,yao2024cacheblend} are incompatible with our experimental setup due to two critical limitations: (1) they require loading the entire model weights into VRAM, and (2) they are not designed to handle the large-scale on-disk knowledge bases central to our evaluation. 
Therefore, we mainly compare our approach against two modern LLM systems with integrated RAG capabilities:
\begin{itemize}[leftmargin=*]
    \item \textbf{RAG with Accelerate (AccRAG).} Hugging Face Accelerate~\cite{HuggingFace} is a well-established system for serving LLMs. It only offloads partial model weights and maintains KV caches on GPU. To ensure sufficient memory for the compute workspace and cache storage, we hereby limit the allocation for weight initialization. 
    \item \textbf{RAG with vLLM (vLLMRAG).} vLLM~\cite{kwon2023efficient} is a state-of-the-art LLM serving system to minimize memory fragmentation and is commonly used in RAG systems~\cite{jin2024ragcache, wang2021milvus}. We utilize vLLM {v0.7.2}, which effectively offloads model weights to RAM and manages computational workspace with the feasible batch size. 
\end{itemize}

\begin{figure}[t]
\centering
\includegraphics[width=1.0\linewidth]{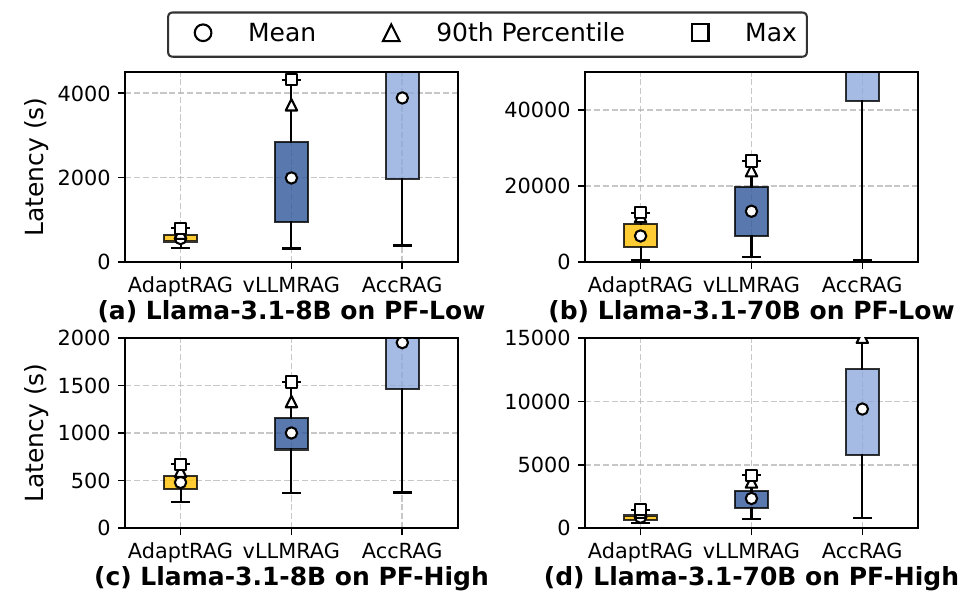}
\caption{Boxplot of all latency values.}
\label{fig:box}
\end{figure}

We integrate the baseline LLM servings with RAG in a serial mode, processing batches in their arrival order with an adaptive size {$4\lambda(t)$} for each interval.
On PF-Low, both baselines encounter memory limitations when serving the 70B model, as the combined GPU and CPU memory is insufficient. To ensure fair comparison, we configure an extended swap space of approximately 180 GB on SSD for PF-Low when integrating the 70B models. To prevent severe performance degradation from excessive I/O operations, we carefully profile and determine the maximum batch sizes that offer optimal throughput and latency performance under memory constrained conditions for the baselines.

\noindentparagraph{Key Metrics.} We target online RAG systems by measuring each request's end-to-end latency, defined as the duration from the request's arrival to its completion. Additionally, we report average latency and various percentile latencies across all requests in selected experiments. Since our proposed systems maintain functional equivalence—preserving the accuracy and quality of outputs without introducing approximations or truncations—our primary focus is on optimizing latency while ensuring consistent result quality.

\begin{figure}[t!]
\centering
\includegraphics[width=1.0\linewidth]{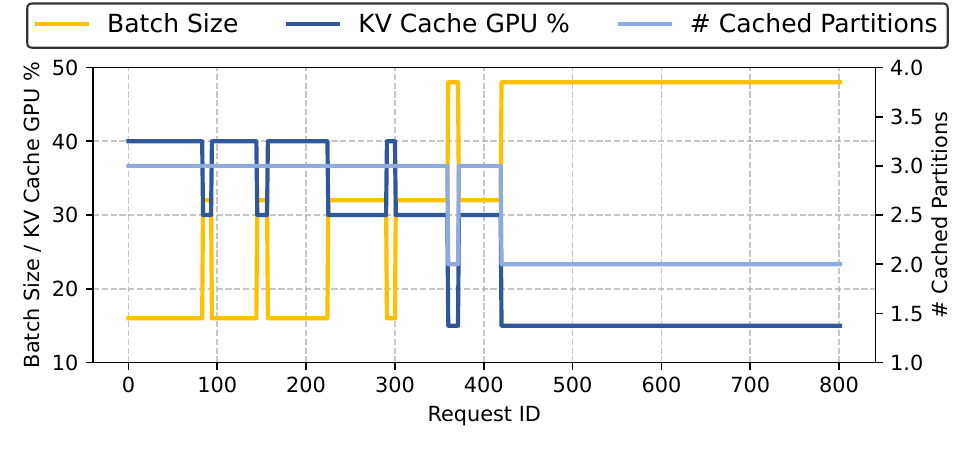}
  \caption{Changes of scheduling of Figure~\ref{fig:perf}(d).}
\label{fig:sche}
\end{figure}

\subsection{End-to-end Performance}
As shown in Figure~\ref{fig:perf}, we report the end-to-end latency of each request during the dynamic workload. {\PP} consistently demonstrates lower average end-to-end latency than the two baselines, achieving a 1.9$\times$ (1236 seconds vs. 2331 seconds in Figure~\ref{fig:perf}(d)) to 3.6$\times$ speedup (555s vs. 1990s in Figure~\ref{fig:perf}(a)) compared to vLLMRAG. Moreover, the speedup exceeds 11.7$\times$ when compared to Accelerate on large models with limited resources (see Figure~\ref{fig:perf}(b)).

For the 8B model, {\PP} maintains an upper bound for each request without experiencing latency explosion under higher workloads. For example, in Figure~\ref{fig:box}(a), every request's latency remains below 1000s, and the trend does not increase as arrival rates rise. In contrast, both baseline methods show increasing latency as arrival rates increase. Although vLLMRAG in Figure~\ref{fig:perf}(c) shows a slower upward trend, its latency still increases from 700-800s (before the 100th request when the arrival rate is 4 requests per minute) to approximately 1500s (as shown in Figure~\ref{fig:perf}(c)) after the 700th request when the arrival rate reaches 16 requests per minute. The performance advantage of {\PP} can be attributed to its more dynamic batch scheduling policy. For instance, in Figure~\ref{fig:perf}(a), {\PP} utilizes notably larger batch sizes, as evidenced by the longer transition segment lengths. Although we also implemented dynamic batch sizing for vLLMRAG, it cannot support the designated larger batch sizes due to limited GPU memory when arrival rates are high. Furthermore, it lacks the ability to pipeline retrieval and generation operations, leading to exponential growth in latency for backlogged requests under limited resources. As for AccRAG, since it does not incorporate prefetching techniques as noted in~\cite{HuggingFace}, its generation efficiency is lower, resulting in easier system overload.

For the 70B model, although all methods encounter  high arrival rates on both machines, {\PP}'s latency escalates more gradually. On PF-Low, where GPU and CPU resources are constrained for 70B model, all methods must use smaller batch sizes, while {\PP} relies on its pipeline architecture and balanced configuration to mitigate the increase. Figure~\ref{fig:box}(b) and (d) show that {\PP} reduces maximum latency by 50\% vs. vLLMRAG and by 80\% vs. AccRAG, demonstrating superior resilience under heavy workloads.

\begin{figure}[t!]
\centering
\includegraphics[width=1.\linewidth]{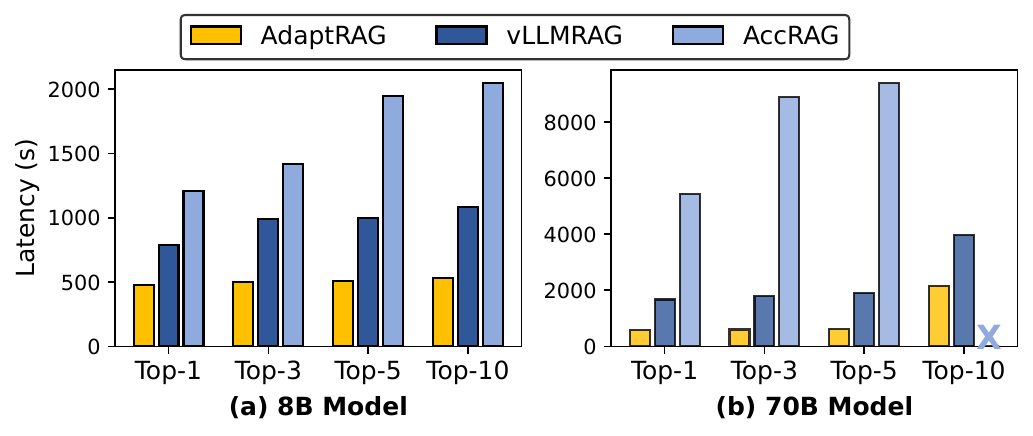}
  \caption{Average latency with different top-k values in retrieval. The “X” mark indicates extreme high latency due to the method’s poor performance.}
\label{fig:top}
\end{figure}

To quantify the adaptive behavior of {\PP}, we analyze the runtime policy adjustments for the 70B model on PF-High as depicted in Figure~\ref{fig:sche}. The data reveals systematic policy transitions as workload intensity increases: the generation batch size progressively scales from 16 to 48, while KV cache CPU allocation decreases from 40\% to 15\%, and vector database partitions cached in CP reduce from 3 to 2. These coordinated policy shifts respond to increasing backlogs under heavier workloads. Following our adaptive scheduling principles, the system prioritizes throughput by expanding batch sizes, which necessitates additional GPU memory allocation. This memory pressure triggers more KV cache offloading to CPU, which in turn requires releasing vector database partitions to disk to maintain memory balance.

The policy selections exhibit some fluctuations during certain phases. Before the 200th request, the generation batch size occasionally varies between 16 and 32 sometimes. These fluctuations arise because {\PP}'s decision making is based on the actual system state rather than simply arrival rate. For instance, when incoming requests are still in the retrieval phase despite high arrival rates, the context queue may remain relatively small, temporarily allowing smaller generation batch sizes. However, as system load stabilizes beyond the 400th request, batch size selection becomes more consistent due to the persistent backlog in the context queue. This behavior demonstrates {\PP}'s adaptive capabilities, enabling efficient resource management without requiring prior knowledge of workload characteristics.

\begin{table}[htbp]
\footnotesize
\setlength{\tabcolsep}{4pt} 
\centering
\caption{Average latency breakdown in seconds.}
\label{tab:breakdown}
\begin{adjustbox}{max width=\columnwidth}
\begin{tabular}{|c|c|c|r|r|r|}
\hline
\textbf{Machine} & \textbf{Model} & \textbf{Mehtod} & \multicolumn{1}{c|}{\textbf{Waiting}} & \multicolumn{1}{c|}{\textbf{Retrieval}} & \multicolumn{1}{c|}{\textbf{Generation}} \\
\hline
\multirow{6}{*}{PF-Low} & \multirow{3}{*}{8B} & {\PP} & 170 & 320 & 66 \\
\cline{3-6}
 &  & vLLMRAG & 1640 & 293 & 57 \\
\cline{3-6}
 &  & AccRAG & 3421 & 288 & 176 \\
\cline{2-6}
 & \multirow{3}{*}{70B} & {\PP} & 5895 & 494 & 466 \\
\cline{3-6}
 &  & vLLMRAG & 12761 & 376 & 222 \\
\cline{3-6}
 &  & AccRAG & 79715 & 357 & 489 \\
\hline
\multirow{6}{*}{PF-High} & \multirow{3}{*}{8B} & {\PP} & 162 & 282 & 36 \\
\cline{3-6}
 &  & vLLMRAG & 677 & 307 & 16 \\
\cline{3-6}
 &  & AccRAG & 1494 & 307 & 151 \\
\cline{2-6}
 & \multirow{3}{*}{70B} & {\PP} & 606 & 388 & 242 \\
\cline{3-6}
 &  & vLLMRAG & 1808 & 303 & 219 \\
\cline{3-6}
 &  & AccRAG & 7936 & 302 & 1152 \\
\hline
\end{tabular}
\end{adjustbox}
\end{table}

\subsection{Latency Breakdown}
 We break down the average latency as shown in Table~\ref{tab:breakdown} to examine each component's contribution. Notably, the waiting time of {\PP} is significantly less than the baselines, while its retrieval and generation times are not necessarily better than vLLMRAG. For instance, with the 8B model on PF-Low, {\PP}'s waiting time is only 10\% of vLLMRAG (170s vs. 1640s), though its retrieval and generation times are slightly higher. This advantage stems from our larger design space and more efficient workload scheduling. Specifically, when workload intensity shifts from low to heavy, we can dynamically arrange larger batch sizes and release more vector database partitions. Although a single request might experience longer retrieval and generation latency, the total latency is reduced through the optimizations analyzed in Section~\ref{sec:tuning}. Additionally, while pipelined retrieval and generation processes may interfere with each other, our pipelined design overlaps substantial latency components, reducing the average latency across all requests. In contrast, vLLMRAG, despite its efficiency in inference through effective swapping optimization, lacks adaptive scheduling capabilities and thus cannot efficiently manage overall requests, resulting in high waiting latency. {\color{black}
Compared to AccRAG, {\PP} reduces waiting and generation times by up to 20$\times$ and 5$\times$, } mainly because AccRAG lacks offloading techniques like prefetching, which leads to significantly longer generation times and accounts for its high overall latency.

\begin{table}[htbp]
    \centering
        \caption{Ablation study of proposed techniques. We report the average latency (seconds) on PF-High. The \textcolor{gray}{gray} number is the static generation batch size policy in the whole process.}
    \label{tab:ablation}
\begin{adjustbox}{max width=\columnwidth}
    \begin{tabular}{l|c|c@{}}
        \hline
        \textbf{Method} & \textbf{8B} & \textbf{70B} \\
        \hline
        {\PP} & 480 & 1236 \\
        Without pipelined design & 663 & 1954 \\
        Without dynamic batch size & 657 \textcolor{gray}{(64)} & 1841 \textcolor{gray}{(32)} \\
        Using FlexGen for LLM inference module & 531 & 1283 \\
        Using vLLM for LLM inference module & 561 & 1432 \\
        \hline
    \end{tabular}
\end{adjustbox}
\end{table}

\subsection{Ablation Study}
To quantify the contribution of each technique in our system, we conduct a comprehensive ablation study. Table~\ref{tab:ablation} presents the latency impact when disabling individual components of {\PP}.
When disabling the pipelined design, we observe the most significant performance degradation: a 38\% increase in latency for the 8B model (663s vs. 480s) and a 58\% increase for the 70B model (1954s vs. 1236s). This substantial impact occurs because, despite retaining other optimizations such as dynamic batch sizing and memory allocation, system flexibility becomes severely constrained. Specifically, the retrieval and generation workers must use identical batch sizes, making it challenging to optimize for both stages simultaneously. Furthermore, both components experience idle waiting periods, reducing overall throughput.

Disabling the dynamic batch size policy also leads to considerable performance deterioration: 657s vs. 480s for the 8B model and 1841s vs. 1236s for the 70B model. This decline stems from the need to pre-allocate memory for the maximum designated batch size, which sacrifices efficiency during periods of low request arrival rates and introduces additional waiting time.

We further evaluate our LLM prefetching module by replacing it with alternatives. First, we integrate a modified version of FlexGen made compatible with Llama models. Despite maintaining all other optimizations, this configuration still results in 10\% and 4\% higher latency than {\PP} for the 8B and 70B models, respectively. The superior performance of {\PP}'s prefetching module is attributed to its pipelined communication and computation design, which maximizes GPU and CPU utilization. The smaller improvement observed for the 70B model occurs because a larger proportion of LLM tensors reside in lower memory hierarchies, making I/O the dominant bottleneck and limiting the additional overlap our prefetching module can achieve.

We also assess the integration of vLLM into our system. While this configuration exhibits significantly better latency than a serial implementation of vLLM in a RAG system (vLLMRAG), achieving 561s vs. 1000s for the 8B model, and 1432s vs. 2331s for the 70B model. However, it still falls short of the performance achieved by {\PP}. Although vLLM can accommodate different batch sizes, it does so within the constraint of fixed weight tensor distribution between GPU and CPU memory. Specifically, while vLLM's interface provides flexibility for a wide range of batch size adjustments, its internal implementation actually determines the effective batch size based on memory constraints rather than directly honoring our specified parameters. Consequently, when batch sizes grow larger, the latency increases almost linearly with batch size, which becomes suboptimal, as discussed in Section~\ref{sec:tuning}. This issue, while present to some degree in both FlexGen and {\PP}, is more pronounced in vLLM. Nevertheless, despite its constrained dynamic capabilities, vLLM can still benefit from our pipelined design and reasonable tuning, enabling it to achieve substantially better performance than vLLMRAG.

In conclusion, we opt for an offloading-based LLM inference system like FlexGen because it offers greater flexibility in dynamically adjusting model weights and KV cache placement across the memory hierarchy. This allows for better trade-offs between throughput and latency under different workloads. However, {\PP}'s implementation can effectively accommodate other types of LLM inference systems such as vLLM, demonstrating the excellent compatibility and extensibility for diverse backend integration.

\begin{figure}[t]
\centering
\includegraphics[width=1.\linewidth]{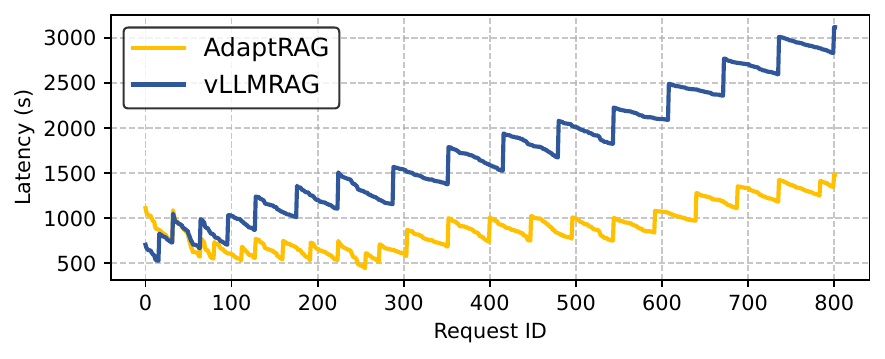}
\caption{Performance of 70B model under DiskANN.}
\label{fig:disk}
\end{figure}

\subsection{Case Study}
\noindentparagraph{Number of Retrieval Chunks.}  We evaluate the impact of different top-k chunks in retrieval. Changes in retrieval chunk count do not significantly affect retrieval latency because loading costs substantially outweigh search costs in the retrieval process, making the impact of additional chunks minimal. However, chunk count substantially impacts generation latency by altering input sequence length. For the 8B model, where generation is not the dominant component, {\PP}'s latency increase remains modest, rising from 480s (for top-1 with input token length under 128) to 529s (for top-10 with input token length under 1024). Similarly, vLLMRAG maintains relative stability (788s to 1080s) because vLLM can maintain the designated batch size for the 8B model on PF-High even as input length approaches 1024 tokens. However, both {\PP} and vLLMRAG experience significant latency growth with top-10 retrieval for larger models, as generation latency becomes dominant. In these scenarios, neither method can utilize larger batch sizes. Nevertheless, {\PP} still maintains a 1.8x speedup over vLLMRAG (2147s vs. 3966s), demonstrating that even when batch size adjustments are constrained, {\PP}'s pipelined design achieves substantially better performance.

\noindentparagraph{On-Disk Index.} 
We also evaluated our method's performance using an on-disk index by creating the DiskANN~\cite{jayaram2019diskann} index of the vector database with Milvus's default configuration. In this setup, searching each partition requires approximately 3.6 GB of CPU memory to store product quantization codes and cache. Consequently, even on PF-High, the system cannot simultaneously accommodate both the index and LLM in memory, necessitating a trade-off in memory allocation. As illustrated in Figure 11, the latency of {\PP} under DiskANN is better than that in the same hardware setup and LLM configuration shown in Figure~\ref{fig:perf}(d) (average 890s vs. 1236s). This improvement is due to {\PP}'s active profiling, which effectively balances the retrieval and generation processes, allowing for reasonable memory allocation between the index and LLM tensors, thereby leveraging DiskANN to enhance retrieval. In contrast, vLLMRAG does not exhibit a noticeable improvement and performs slightly worse (2427s vs. 2331s) under the same setup. This is because, although the DiskANN index saves more memory than the original data, reloading the index for each partition is more time-consuming than loading the original data due to additional disk I/O operations and index initialization overhead. To fully utilize the on-disk index, reserving more memory for the index is necessary, which may impact the efficiency of LLM inference. This underscores the complexity of optimizing scheduling and allocation policies in RAG systems under resource constraints. In summary, while the patterns differ between using and not using an on-disk index for retrieval, we believe this can be viewed as a shift in trade-offs between retrieval and generation, allowing {\PP} to optimize accordingly. Moreover, employing an on-disk index can enhance the scalability of the system by enabling efficient processing of much larger knowledge bases. This demonstrates the practical value of our approach for deployment in realistic production environments.

\section{Related Work}

\noindentparagraph{Efficient LLM Serving.} 
To enable efficient LLM inference under strict performance and resource constraints, numerous systems~\cite{raistrick2024infinigen,zheng2024sglang,eliseev2023fast,du2025flexinfer,wu2024loongserve,aminabadi2022deepspeed,tan2024pipellm,jiang2024neo} improve inference efficiency by offloading KV caches or model weights to limit GPU memory usage. 
{A line of research~\cite{sheng2023flexgen,song2024powerinfer,zhao2024hetegen} represented by FlexGen prioritize inference throughput. In comparison, systems like vLLM~\cite{kwon2023efficient} particularly enhance online latency through batch scheduling techniques.}
{\PP} is compatible with many of these optimization strategies. In this paper, we adopt FlexGen as our representative implementation due to its flexible tensor placement capabilities, which align well with our focus on resource-limited settings.
Another approach to reduce LLM size is model compression~\cite{dettmers2022gpt3,frantar2022gptq,yao2022zeroquant,xiao2023smoothquant}, since it does not inherently address the memory issue, it is commonly regarded as an auxiliary solution alongside offloading methods in practical deployments.

\noindentparagraph{LLM Computation for RAG.}
RAG~\cite{lewis2020retrieval,ram2023context,li2022survey,gao2023retrieval,mao2020generation} enhances LLM generation by incorporating external knowledge. Recent studies ~\cite{jin2024ragcache,jiang2024piperag,zhu2024accelerating,yao2024cacheblend,lu2024turborag,jiang2025rago,jiang2023chameleon} dedicate to more efficient LLM serving tailored for RAG, primarily focusing on the management of KV cache inside generation computation. 
\PP{} is complementary to these techniques with the capability in jointly optimizing retrieval and generation under both computation and storage constraints. 

\noindentparagraph{On-disk Vector Retrieval.} 
Production-grade vector databases \cite{Milvus,Weaviate,Chroma,pinecone} offer fundamental retrieval capabilities for RAG systems. 
Indexing algorithms~\cite{jayaram2019diskann,gollapudi2023filtered,singh2021freshdiskann} further power data retrieval by constructing on-disk indices to boost query lookup and reduce memory overhead. 
{\PP} complements these approaches by systematically optimizing their efficiency under resource-constrained RAG deployments.

\section{Conclusion}
We present {\PP}, a self-adaptive LLM-RAG integration on a consumer-grade GPU. By implementing joint memory management, enhanced multi-pipeline processing, and adaptive batch scheduling, our system achieves up to 3.6$\times$ speedup compared to state-of-the-art baselines. These results demonstrate that carefully designed resource allocation strategies can make large-scale RAG practical on resource-constrained devices.


\bibliographystyle{ACM-Reference-Format}
\bibliography{sample-base}

\newpage
\mbox{}
\newpage

\end{document}